\documentclass[conference]{IEEEtran}

\usepackage{epsfig} 
\usepackage{epstopdf}
\usepackage{amsmath} 
\usepackage{graphicx}
\usepackage{amsfonts}
\usepackage{float}
\usepackage{color}
\usepackage{cite}
\usepackage{booktabs,subcaption,dcolumn}
\usepackage{url}
\usepackage{hyperref}
\usepackage{array}
\usepackage{multirow}
\usepackage{multicol}
\usepackage{caption}
\newcolumntype{P}[1]{>{\centering\arraybackslash}p{#1}}

\usepackage{lipsum}

\newcommand\blfootnote[1]{%
  \begingroup
  \renewcommand\thefootnote{}\footnote{#1}%
  \addtocounter{footnote}{-1}%
  \endgroup
}

\title{Millimeter-Wave V2X Channels: Propagation Statistics, Beamforming, and Blockage}
\author{Chethan Kumar Anjinappa and Ismail Guvenc \\
\thanks{*This work was not supported by any organization}
\IEEEauthorblockA{Department of Electrical and Computer Engineering, North Carolina State University, Raleigh, NC}
Email: {\tt \{canjina,iguvenc\}@ncsu.edu}
\vspace{-0.3cm}
}

\begin{document}

\maketitle
\thispagestyle{empty}
%
\pagestyle{empty}


\begin{abstract}
5G millimeter wave (mmWave) technology is envisioned to be an integral part of next-generation vehicle-to-everything (V2X) networks and autonomous vehicles due to its broad bandwidth, wide field of view sensing, and precise localization capabilities. The reliability of mmWave links may be compromised due to difficulties in beam alignment for mobile channels and due to blocking effects between a mmWave transmitter and a receiver. 
In this paper, we study the channel characteristics for mmWave and sub-6 GHz V2X communications using ray-tracing simulations. We present results for time-varying path-loss, delay spread, and angular spreads in the presence of a moving vehicle for line-of-sight (LOS) and non-LOS (NLOS) trajectory, respectively. Additionally, we study the effect of delay spread and angular spread when the base station (BS) and user equipment (UE) are capable of adjusting the beam directions and beamwidths dynamically. Finally, we explore the impact of blockage effects on the quality of receive beams at 28 GHz for the considered vehicle trajectories.
\blfootnote{This research was supported in part by NSF under ACI-1541108.}
\end{abstract}

\begin{IEEEkeywords}
 5G, autonomous vehicles, mmWave, ray tracing, side information, sparsity, V2I, vehicular communication.  
\end{IEEEkeywords}

\section{Introduction}

Millimeter wave (mmWave) is envisioned as a promising frequency band for the next generation mobile, vehicular-to-everything (V2X) broadband wireless networks due to large spectrum resources available. It has a great potential in realizing data rates beyond gigabits-per-second, significantly exceeding what is possible with traditional cellular systems operating at sub-6~GHz bands. However, communication at mmWave bands, especially for vehicular scenarios, is challenging because of the severe path-loss (PL) and the Doppler effects due to smaller wavelengths at mmWave frequencies. 

To establish a strong link that conveys sufficient signal power, it is necessary to use directional beam-forming at the mmWave base station (BS) and/or user equipment (UE). In addition to improving the signal strength, directional reception can also increase the channel coherence time~\cite{Temporal_Variation_Beamwidth}. 
However, for the directional transmission/reception the best transmit and receive angles needs to be known between the BS and the UE, in order to establish a strong link that conveys sufficient signal power. Finding such directions is referred to as the beam alignment (BA) problem \cite{BA_Problem} which is challenging and is often attained at the expense of BA overhead. 
%
From recent works \cite{7888146,8198818,ICC_Workshop}, it is known that there exists strong congruency in the temporal and angular domain across sub-6~GHz and mmWave bands. Thus, one can exploit spatial information obtained from sub-6~GHz to find the best transmit/receive directions and beamwidths (BeWs) dynamically at the mmWave bands. 

An important parameter in the use of directional antennas is the root-mean-square (RMS) directional angular spread (AS) and the delay spread (DS) over time. In the past, considerable research is done to characterize the vehicular channel models sub-6 GHz and mmWave bands. For example, \cite{5GHz_Matolak} presents the vehicle-to-vehicle channel models and its characteristics in the 5~GHz band. The channel characteristics at 28 GHz and 38 GHz are studied in \cite{28GHz} and \cite{38GHz}, respectively, while \cite{60GHz_Rappaport,60GHz_Kato} presents the propagation characteristics of inter-vehicle communication at 60 GHz for stationary vehicles.  
Most of these work use dedicated hardware which are often expensive and requires time-consuming efforts to obtain measurements. 

In this work, we use ray-tracing (RT) simulations, which have been reasonably successful in the literature at predicting site-specific propagation characteristics at sub-6 GHz and mmWave bands, if an accurate description of the measurement environment is used in simulations. 
To the best of our knowledge, there exists no work in the literature that studies the effect of sub-6 GHz and mmWave bands together, which necessitates how propagation characteristics are coupled across different bands. 
In this work, we investigate the temporal/angular channel characteristics for few of the popular sub-6~GHz and mmWave bands with and without beam-forming. In addition, we study blockage effects on the beamforming link quality for V2X propagation channels at 28 GHz. 

The rest of this paper is organized as follows. In Section~II, we present the system model and RT simulation setup. Results for average PL, RMS-AS, and RMS-DS, as observed at the BS and UE in the presence of moving vehicles, are presented in Section III and Section IV, respectively. Additionally, we investigate the effects on RMS-AS and RMS-DS at mmWave bands when the BS and UE are capable of adjusting the beam directions and BeWs dynamically. In the same section, considering blockage effects, we investigate the angular characteristics and power concentration of the multipath components with dynamic BA. Finally, we conclude the paper in Section~V.

\section{System Model and Ray-tracing Simulations}\label{System_Model}


\subsection{System Model}
Consider the following physical propagation model between the transmitter and receiver through the omni-directional channel impulse response (CIR):
\begin{equation}\label{DCIR1}
h(t,\boldsymbol{\tau},\boldsymbol{\psi},\boldsymbol{\theta}) = \sum_{k=1}^{\tilde{K}(t)} \underbrace{\rho_k e^{-j \phi_k}}_\text{AnP term} \underbrace{\delta (\tau - \tau_k)}_\text{TOA term} \underbrace{\delta (\psi - \psi_k)}_\text{AOA term} \underbrace{\delta (\theta - \theta_k)}_\text{AOD term}
\end{equation}
where $t$ is the time at which the channel is excited, $\tau_k$ is the considered time delay, $\psi_k$ and $\theta_k$ are the angle-of-arrival (AOA) and angle-of-departure (AOD) associated with the $k^\text{th}$ multi-path component (MPC), respectively. The term $\rho_k$ is the amplitude which is based on the PL and shadowing associated with the $k^\text{th}$ MPC and $\phi_k$ is the phase component associated with the $k^\text{th}$ MPC which primarily depends on the time delay and the Doppler shift. All the TOA, AOA, and AOD components are further clustered into the vectors $\boldsymbol{\tau}=[\tau_1,...,\tau_{\tilde{K}(t)}]$, 
$\boldsymbol{\psi}=[\psi_1,...,\psi_{\tilde{K}(t)}]$, and $\boldsymbol{\theta}=[\theta_1,...,\theta_{\tilde{K}(t)}]$, respectively. The amplitude and phase (AnP) terms are merged under the AnP variable, $\boldsymbol{\gamma}~=[\rho_1 \exp(-j\phi_1),...,\rho_{\tilde{K}(t)} \exp(-j\phi_{\tilde{K}(t)})]$. Intuitively, (\ref{DCIR1}) is the sum of the contributions of discrete MPCs, where, the number of MPCs $\tilde{K}(t)$ may itself be time-varying due to the mobility of the vehicle and the surrounding scatterers. Note that directional propagation channel can be extracted from~\eqref{DCIR1} by exciting subset of the MPCs, weighted by transmit/receive beam shapes in the angular domain. 

\begin{figure}[b!]
\includegraphics[scale=0.435]{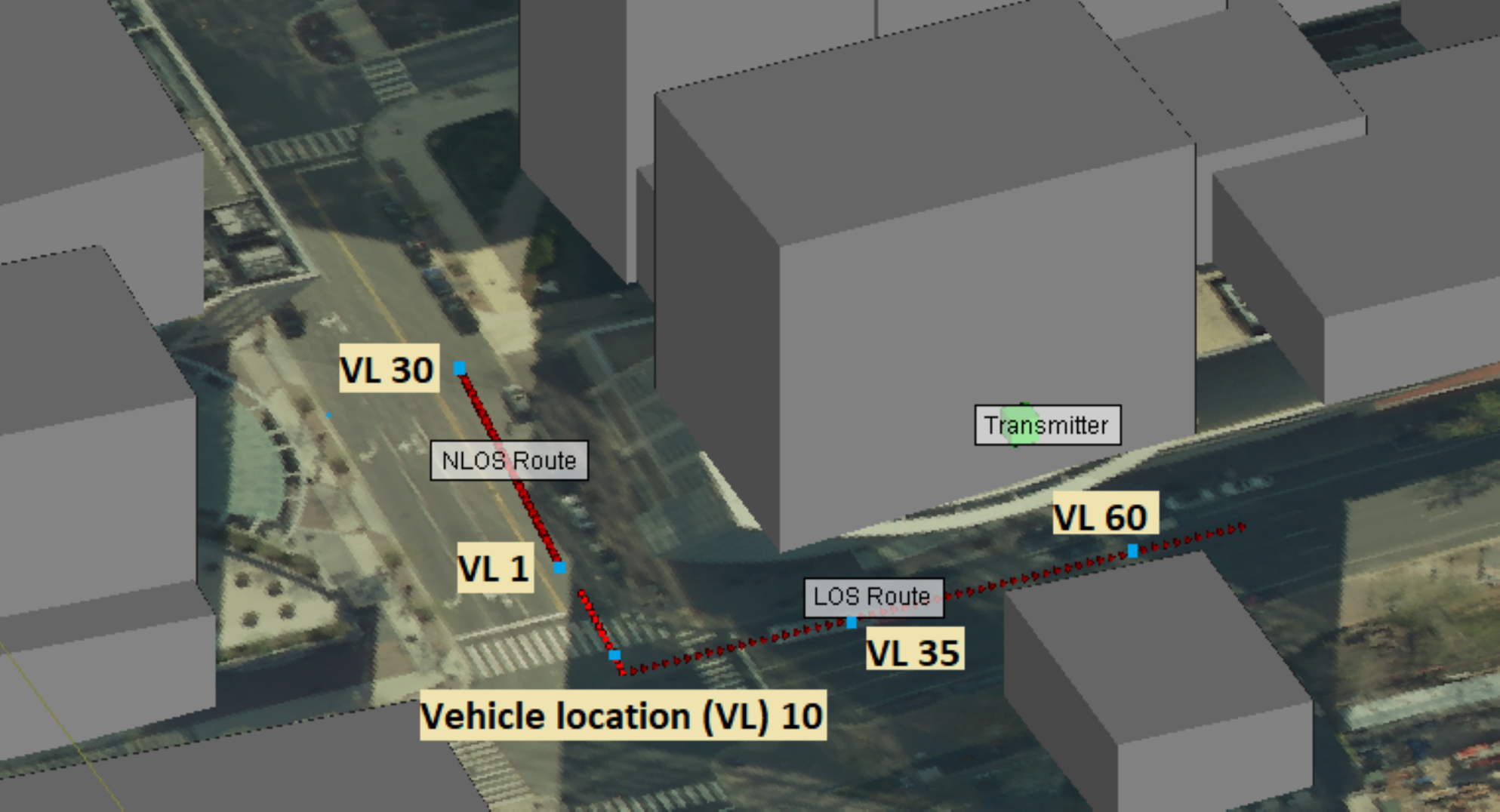}
\caption{Side view: Vehicular ray-tracing scenario in North-Moore Street, Rosslyn, Virginia.}
\label{Ray_Tracing1}\vspace{-.5cm}
\end{figure}

\subsection{Ray-tracing Simulations}\label{Sec:RT}
In this work a full 3D RT simulator, WirelessInsite~\cite{WirelessInsite}, is used as a tool for the analysis of site-specific radio wave propagation. 
In the past, RT has been reasonably successful at predicting site-specific mmWave propagation~\cite{Comparison_RayTracing}. The mmWave frequencies can be well approximated by ray optics approaches as the diffraction is low and scattering is mostly limited to surface reflections~\cite{Air_Interface_Ray_Tracing_Study}. Hence, with proper approximations, RT simulators can accurately model the mmWave and sub-6~GHz channels. We consider a vehicle-to-infrastructure (V2I) scenario in an urban environment where a moving vehicle communicates with an infrastructure (cellular BS) as in Fig.~\ref{Ray_Tracing1}. We set~up a vehicular UE (red path) and a BS node (green point). All the nodes use half-wave dipole antennas that are excited with varying frequencies that are shown in Table~\ref{tab:T_FI}. For simplicity, single input single output (SISO) model is considered in this work, that is, we assume both the transmitter and receiver is equipped with a single antenna.

\begin{table}[t!]
\caption{Frequency bands used in ray tracing simulations.}
\begin{tabular}{|P{2.05cm}|P{0.325cm}|P{0.325cm}|P{0.325cm}|P{0.325cm}|P{0.325cm}|P{0.325cm}|P{0.325cm}|P{0.325cm}|}
\hline
\textbf{Band} & \textbf{$f_1$} &  \textbf{$f_2$} &  \textbf{$f_3$} &  {\textbf{$f_4$}} &  \textbf{$f_5$} & {\textbf{$f_6$}} &  \textbf{$f_7$} &  \textbf{$f_8$}\\
\hline
\textbf{Frequency(GHz)}  & 0.9& 1.8 & 2.1 &  {5.0} & 5.9& {28} & 38 & 73\\
\hline
\end{tabular}\label{tab:T_FI}\vspace{-.45cm}
\end{table}

The BS antenna in the RT simulations is placed at a height of 10~m from the ground and transmits at a power level of 0~dBm. The moving UE antenna is placed at a height of 2~m from the ground and takes the indicated route (red path) in Fig.~\ref{Ray_Tracing1}. There exists a total of 100 vehicle locations (points) in the route: the first 30 points correspond to the NLOS trajectory and the latter 70 points correspond to the LOS trajectory. The rationale behind choosing separate LOS and NLOS paths is that the mmWave channel is often a LOS or quasi-LOS channel with few dominant path clusters.  

At mmWave frequencies, surface roughness impacts wave propagation, causing scattering in non-specular directions that can lead to a significant effect on received signal strength and polarization. To accurately predict channel characteristics for mmWave frequency bands, propagation modeling must account for diffuse scattering effects. To include the effect of diffuse scattering, we enabled the option of ``$\textit{diffuse scattering}$" provided by the WirelessInsite and set the maximum number of reflections, transmissions, and diffractions to 6, 0, and 1, respectively. With these settings, we carry out the RT simulations for some of the popular sub-6 GHz and the mmWave bands as shown in Table~\ref{tab:T_FI}. From RT simulations, the vectors $\boldsymbol{\tau}$, $\boldsymbol{\theta}$, $\boldsymbol{\psi}$ and $\boldsymbol{\gamma}$ for a total of 250 MPCs, $\boldsymbol{\tilde{K}}(t)$, were collected for each vehicle location.\footnote{Data available online: \href{https://research.ece.ncsu.edu/mpact/data-management/}{MPACT repository: V2I RT data.}}


\section{Channel Statistics Based on RT simulation}
In this and the subsequent section, we demonstrate the simulation results relating to the temporal and angular characteristics at the sub-6~GHz and the mmWave bands.

\subsection{Impact of Number of Reflections per Path}
\begin{figure*}[t!]	\centerline{{\hspace{-0.4cm}\includegraphics[scale=0.425]{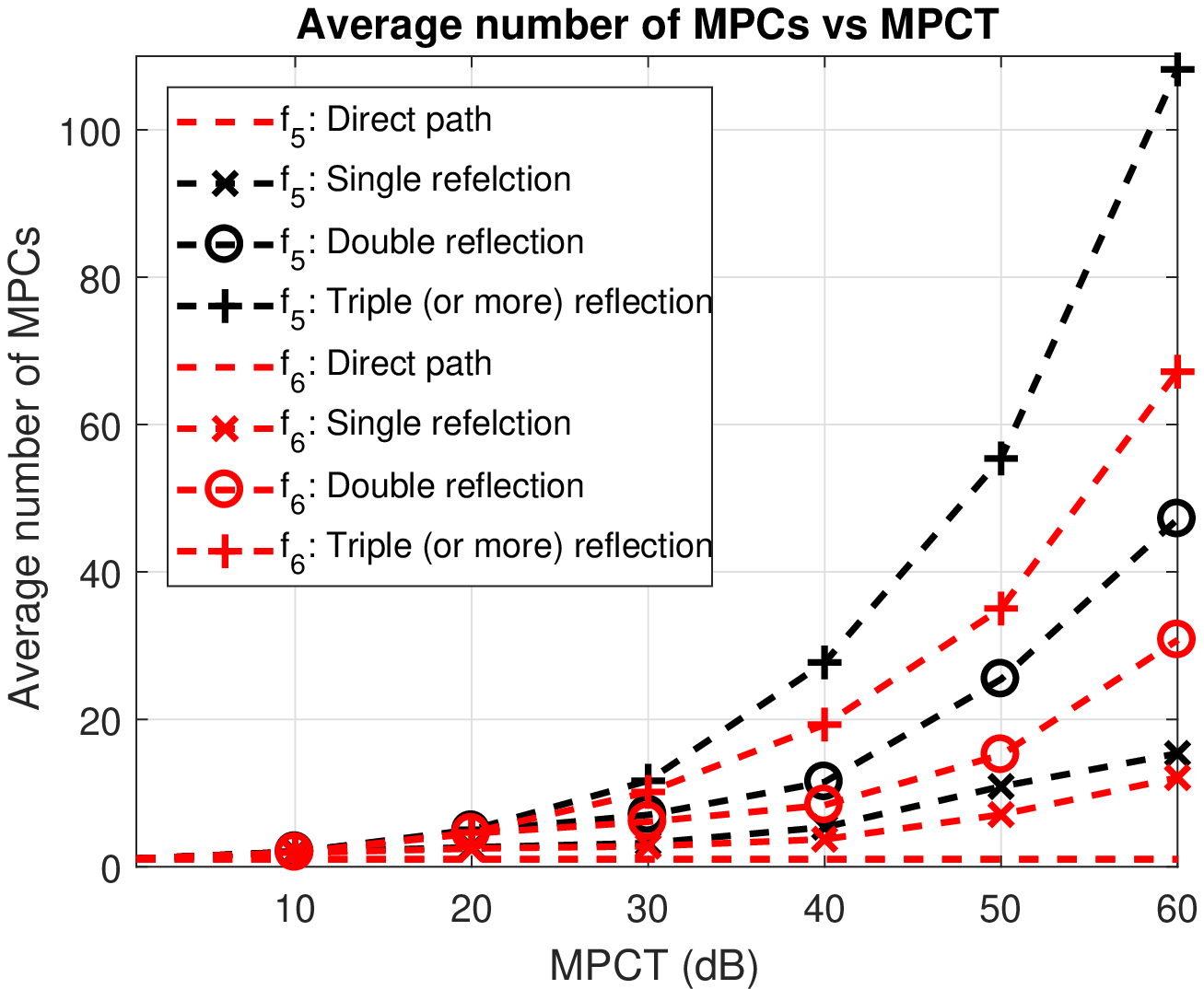} \hspace{-0.4cm} \includegraphics[scale=0.425]{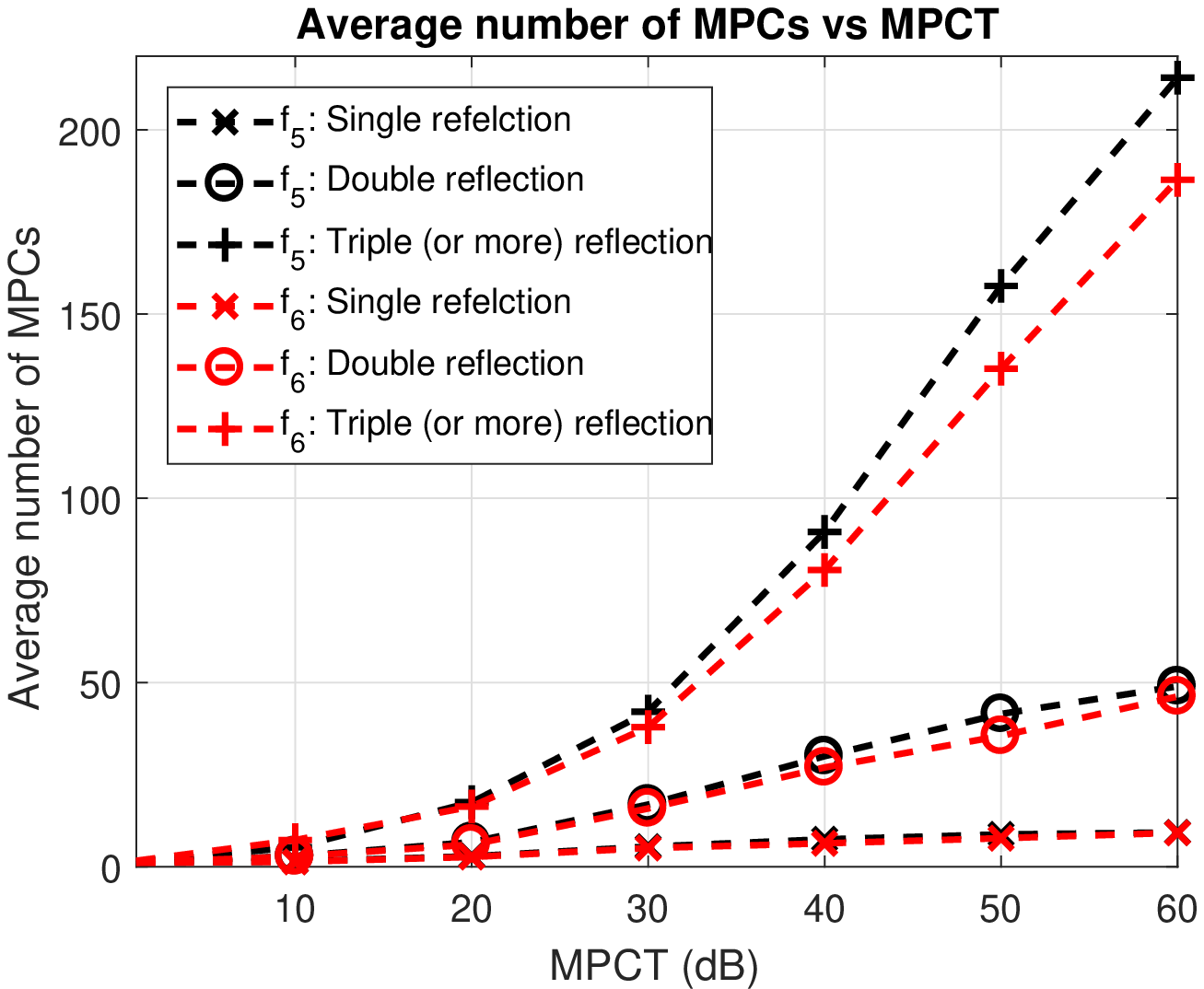} \hspace{-0.4cm}\includegraphics[width=6cm,height=4.75cm]{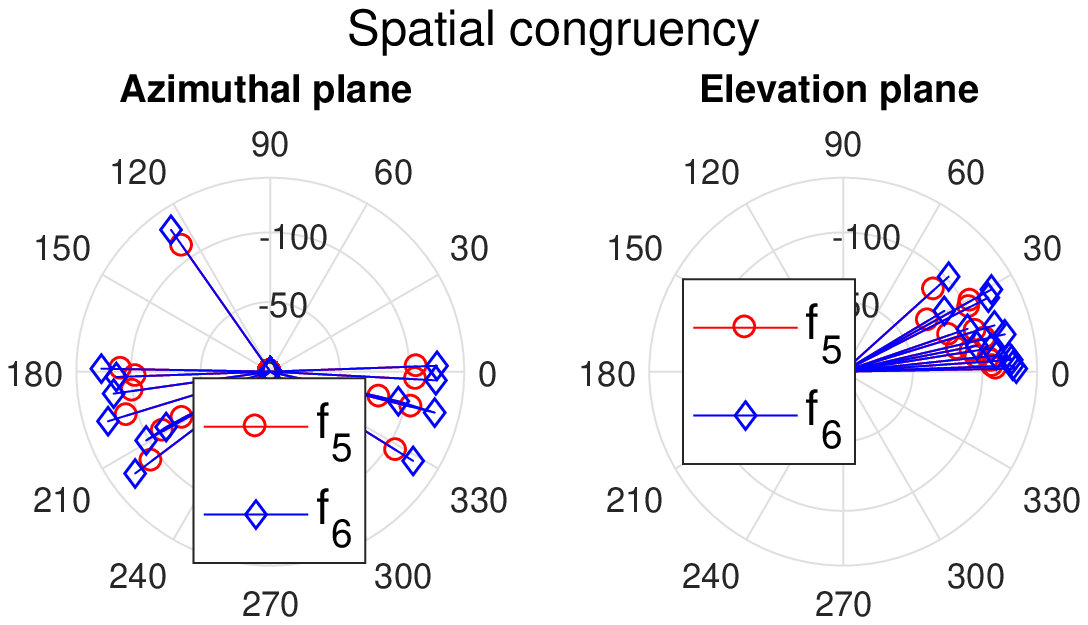}}}
            \caption{Average number of MPCs captured as a function of the MPCT value for bands $f_5$ and $f_6$. \textbf{Left:} LOS trajectory, \textbf{Middle:} NLOS trajectory, and \textbf{Right:} AOA at $f_5$ and $f_6$ for the vehicle location 60 (LOS trajectory) with MPCT = 40 dB (Left: AOA at the azimuth plane, Right: AOA at the elevation plane).}\vspace{-.55cm}
            \label{Average_MPCs_and_Polar_Plot}
\end{figure*}

\begin{figure}[t!]
	\centerline{\includegraphics[scale=0.35]{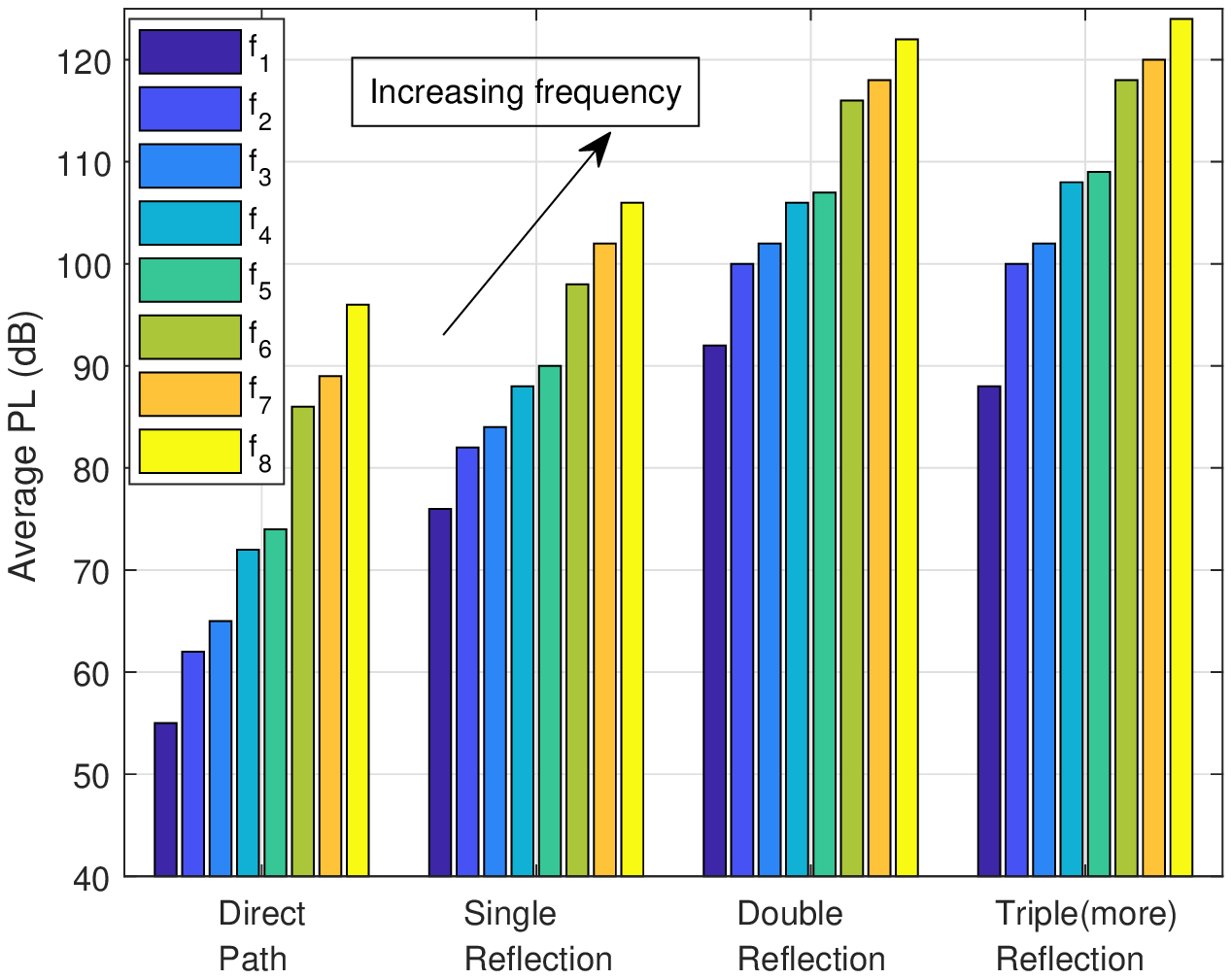}\hspace{-0.65cm}\includegraphics[scale=0.35]{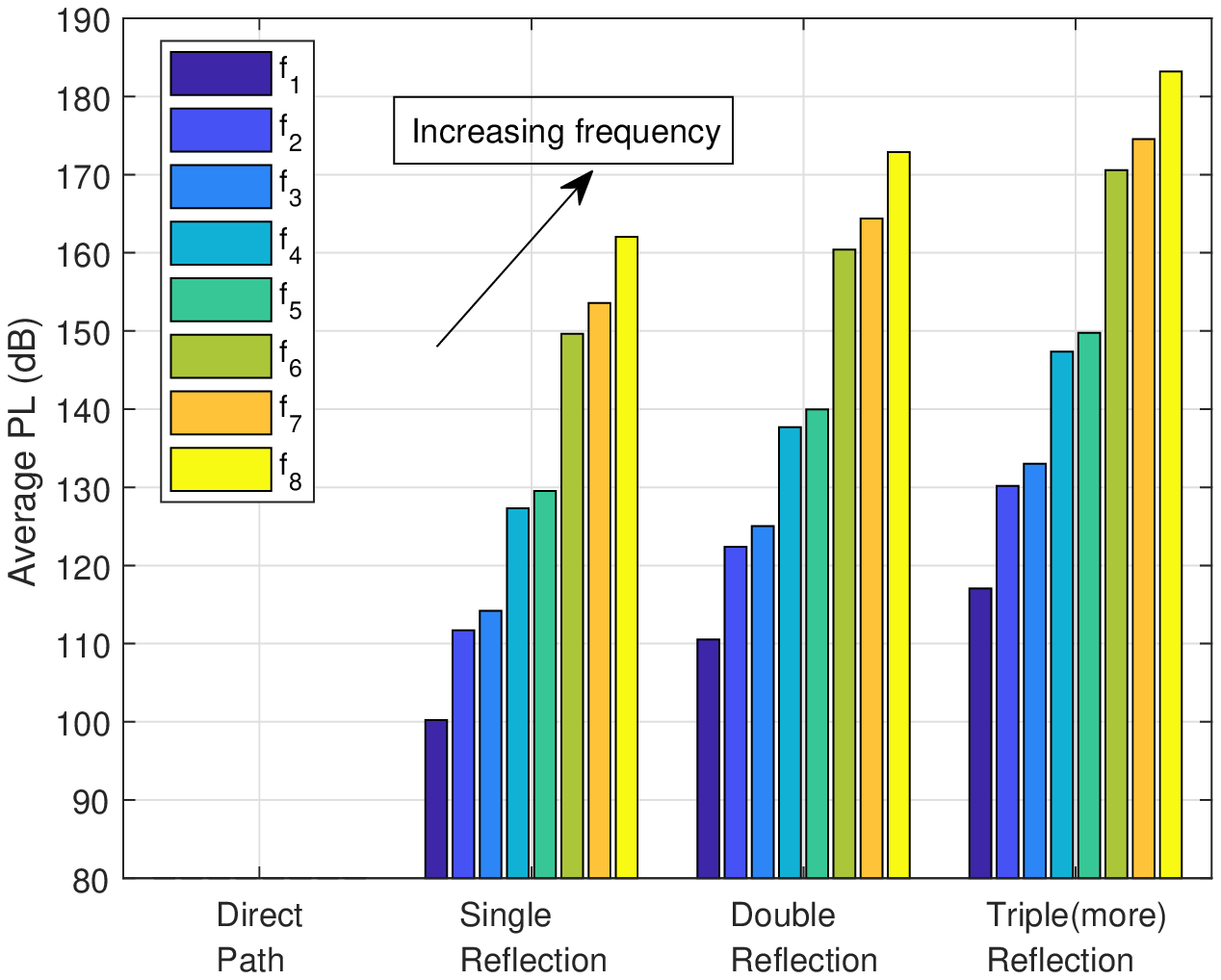}}
    \caption{Average PL for direct path and multiple bounce reflections with MPCT = 40 dB. \textbf{Left:} LOS Trajectory and \textbf{Right:} NLOS trajectory.}
    \label{Avg_PL_Multiple_Bounce}\vspace{-.5cm}
\end{figure}

The RT simulation results include 250 rays (MPCs) per link, many of which would often go undetected in the actual measurement equipment due to the limited dynamic range of the equipment. 
One can limit the number of MPCs based on a thresholding approach: we can consider the MPCs to be present if and only if the MPC strength is above a threshold relative to the most dominant MPC. We refer to this threshold as our ``MPC Threshold (MPCT)". The choice of MPCT depends on the trade-off between an accurate representation of the channel and the implementation complexity. Thus, there is a necessity of choosing a right MPCT value before we perform any data processing on the collected data.

For the sake of deciding on the MPCT value, we plot the average number of MPCs captured as a function of different MPCT values. Fig.~\ref{Average_MPCs_and_Polar_Plot} plots the above mentioned effect for bands $f_5$ and $f_6$, which are the two popular bands for V2X communication, for both the LOS and NLOS trajectory, respectively. The plot also considers the distribution of the number of MPCs including the direct path, single and double reflections, and beyond double reflections. 
This provides an insight on the interaction of MPCs with the environment.

As evident from the plots, the average number of MPCs is smaller in the mmWave band $f_6$ compared to sub-6 GHz band $f_5$. Having an MPCT of 40 dB would capture MPCs that have significant energy in both the LOS and NLOS trajectory, key statistics discussed in \cite{ICC_Workshop}. Often for the LOS trajectory, the direct path, the single and double reflections carry significant energy, whereas, for the NLOS trajectory only the single reflections are the dominant ones. As the MPCT is increased, more and more MPCs are included; however, these MPCs undergo multiple-bounce reflections (double and beyond double reflections) and often carry insignificant energy which can be neglected.
 
Based on the above discussion, we set the MPCT value to 40 dB for both the LOS and NLOS trajectory. Before we perform any data processing on the collected data, we apply an MPCT of 40 dB on the collected data and process it further. We note that channel parameters extracted from the data can be significantly different when using a different MPCT value. Fig. \ref{Avg_PL_Multiple_Bounce} plots the average PL for MPCT = 40 dB at different frequency bands for both the LOS and NLOS trajectory, respectively. The plot shows the 
average PL for direct and multiple-bounce reflections, which increases significantly for larger number of reflections at all bands.


  
\subsection{Spatial Congruency}
In this sub-section, we study 
the spatial congruency across the sub-6 GHz and mmWave bands. The spatial congruency and correlation across multiple bands have been explored in detail in our previous work \cite{ICC_Workshop}. In here, for simplicity, we show spatial congruency at bands $f_5$ and $f_6$. Fig.~\ref{Average_MPCs_and_Polar_Plot} shows the congruency in the AOA domain; the radius value corresponds to the received power in dBm. Lower the radius of the MPC, more the received power. As evident, the overlap in the angular domain is significant albeit not perfect. Thus, the spatial information from lower band can be used to perform BA and selection of BeWs at the mmWave bands.

The key point from these results is that there is a significant spatial congruency across the mmWave and sub-6~GHz bands. Thus, as a general comment, if the spatial information from the out-of-band (OOB) is available then it is more sensible to exploit it for the benefit of robust and efficient communication. The results in here do not take into account the effects of blockage, small reflectors at mmWave frequencies, which we will explore further through RT simulations in our future work.

\begin{figure*}[h!]
    \centering
    \begin{subfigure}[t]{0.5\textwidth}
			\centerline{\includegraphics[scale=0.425]{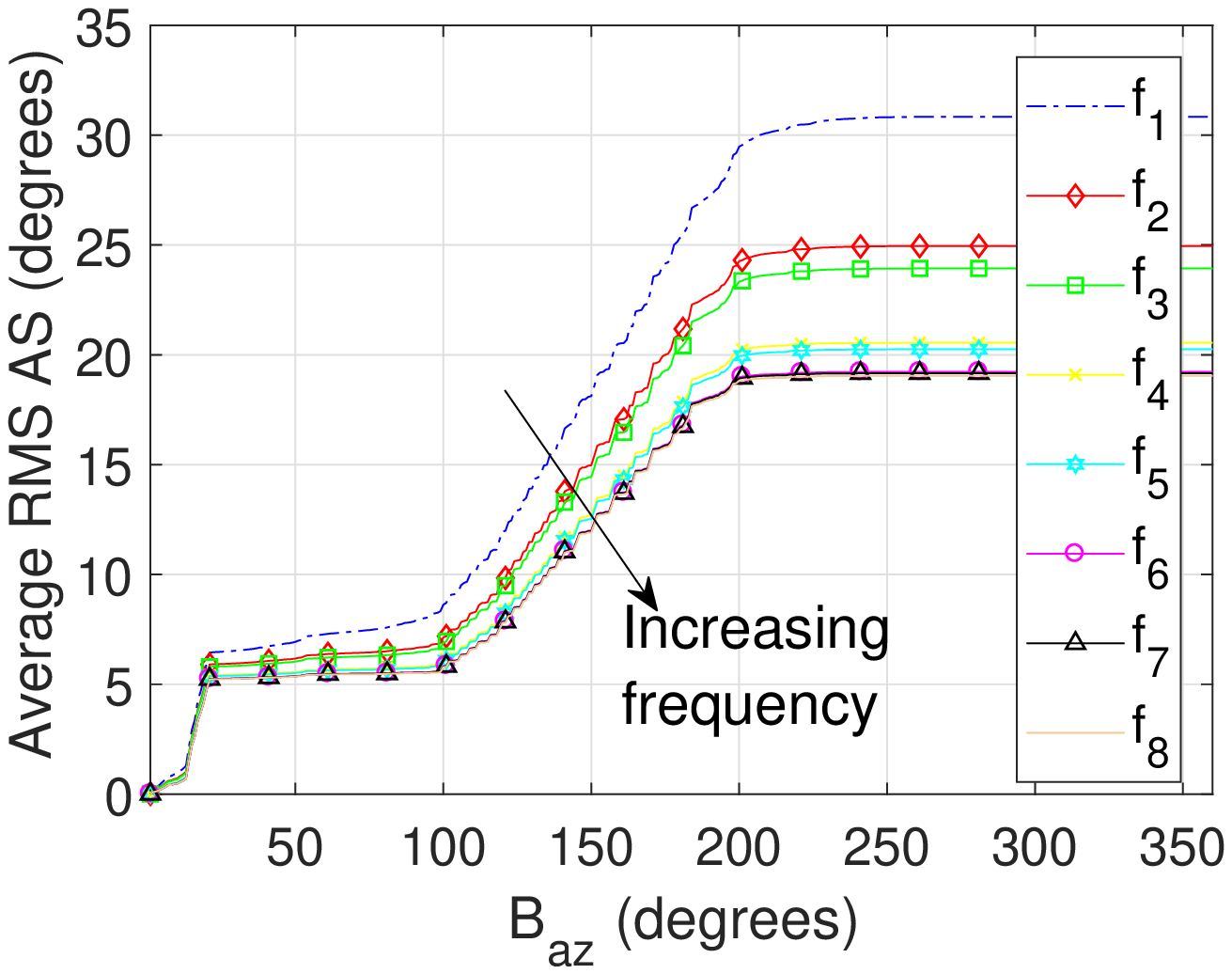}\includegraphics[scale=0.425]{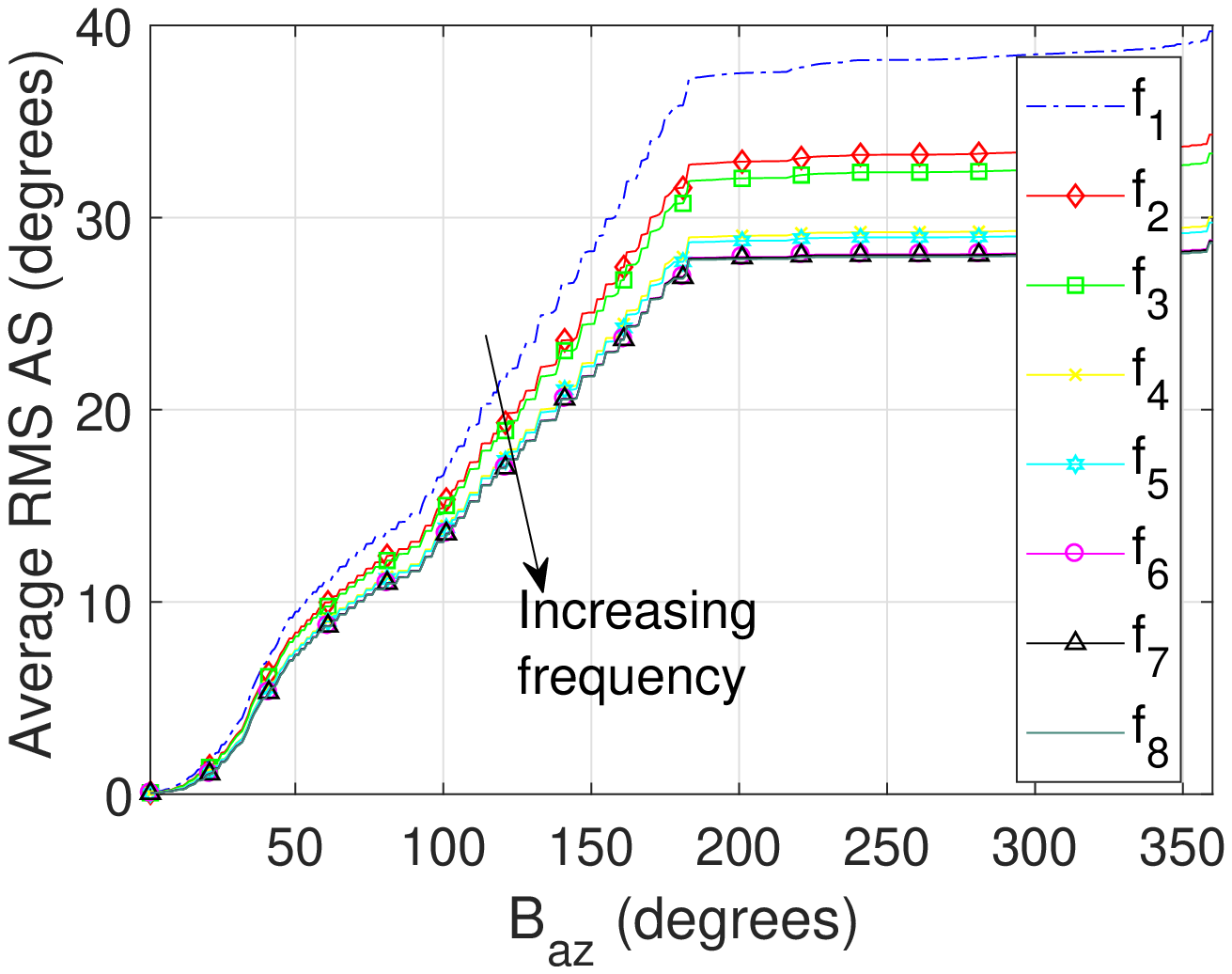}\includegraphics[scale=0.425]{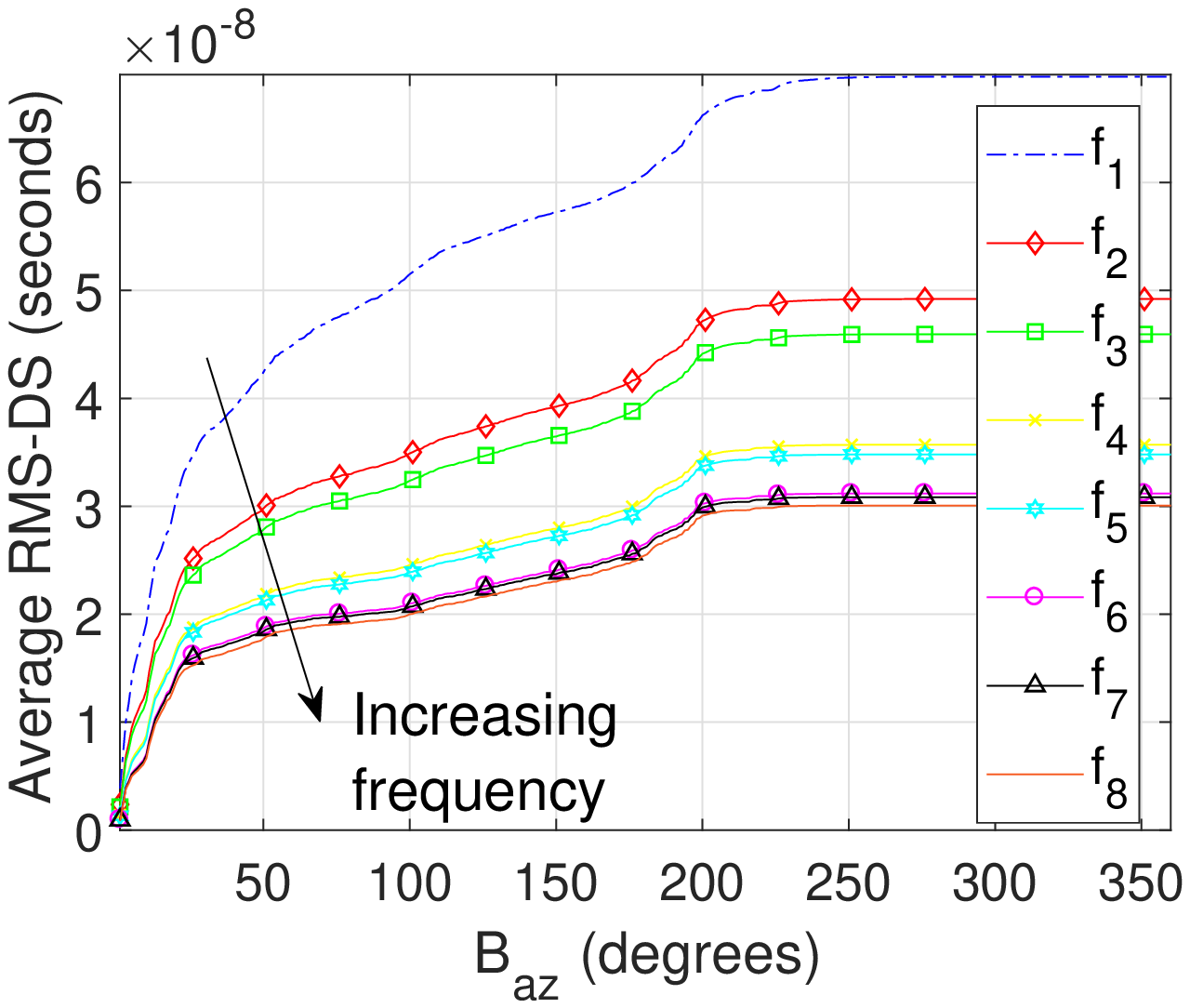}}
            \caption{Azimuthal plane.}
    \end{subfigure}%
        \vspace{0.05cm}
    \begin{subfigure}[t]{0.5\textwidth}
       \centerline{\includegraphics[scale=0.425]{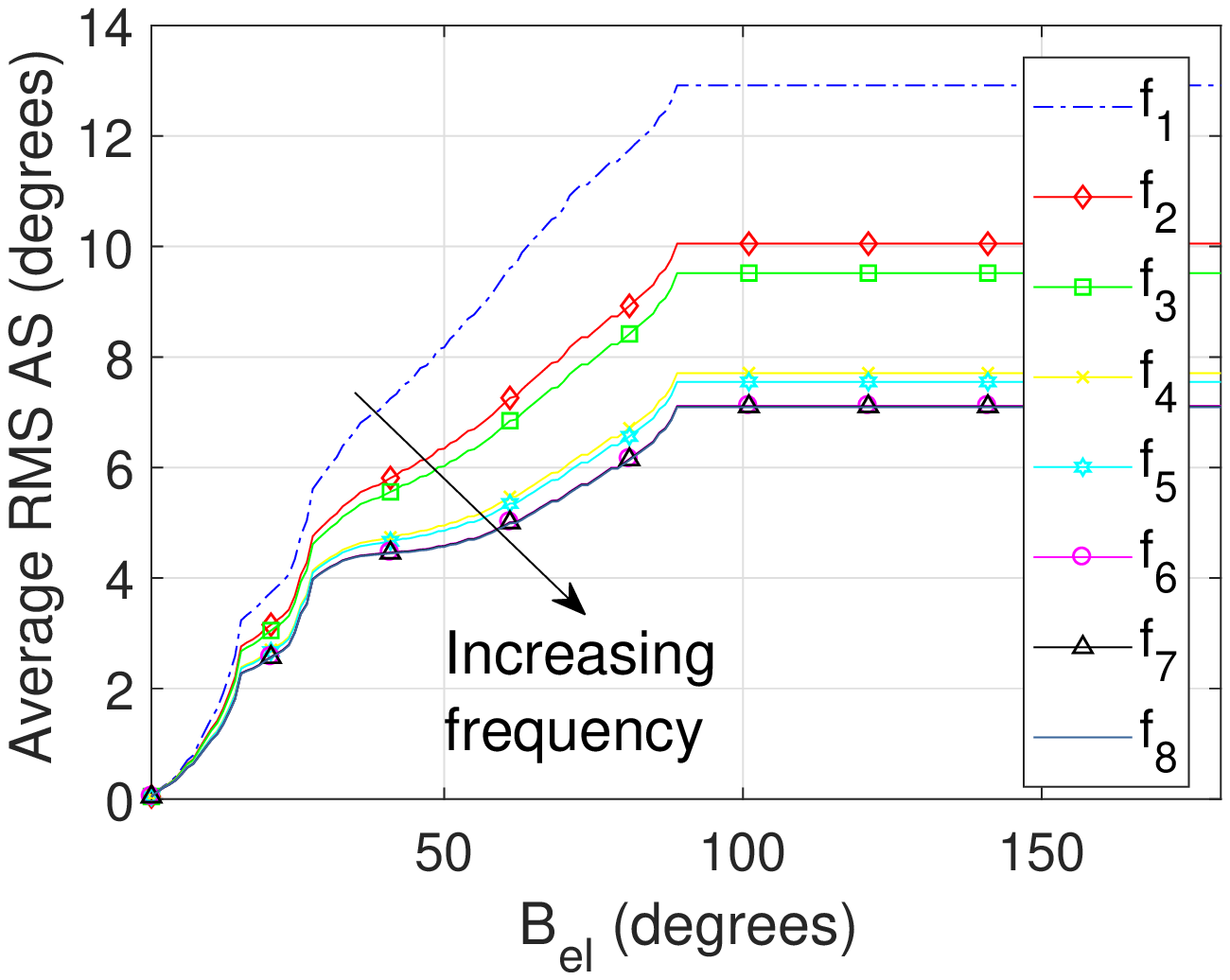}\includegraphics[scale=0.425]{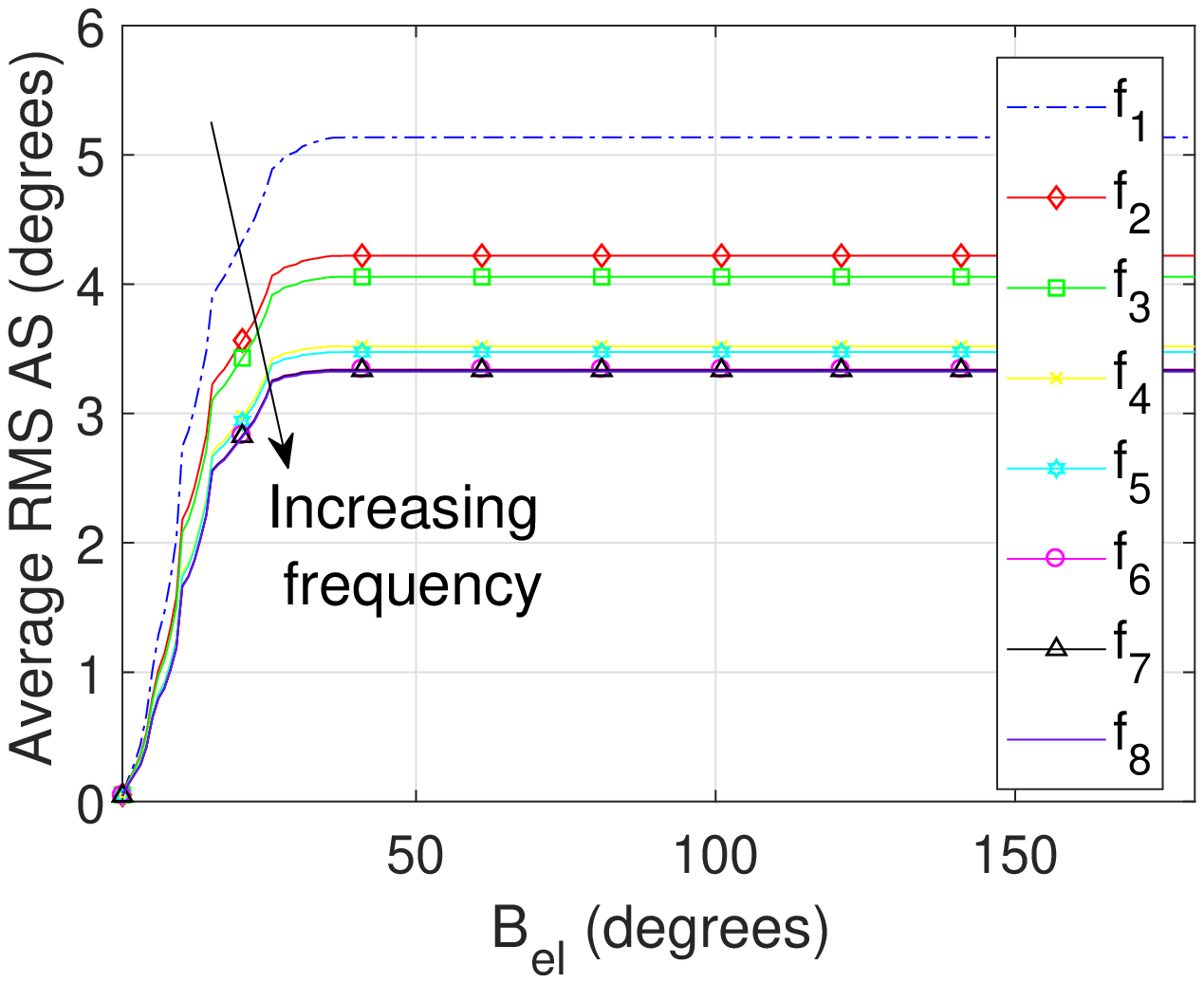}\includegraphics[scale=0.425]{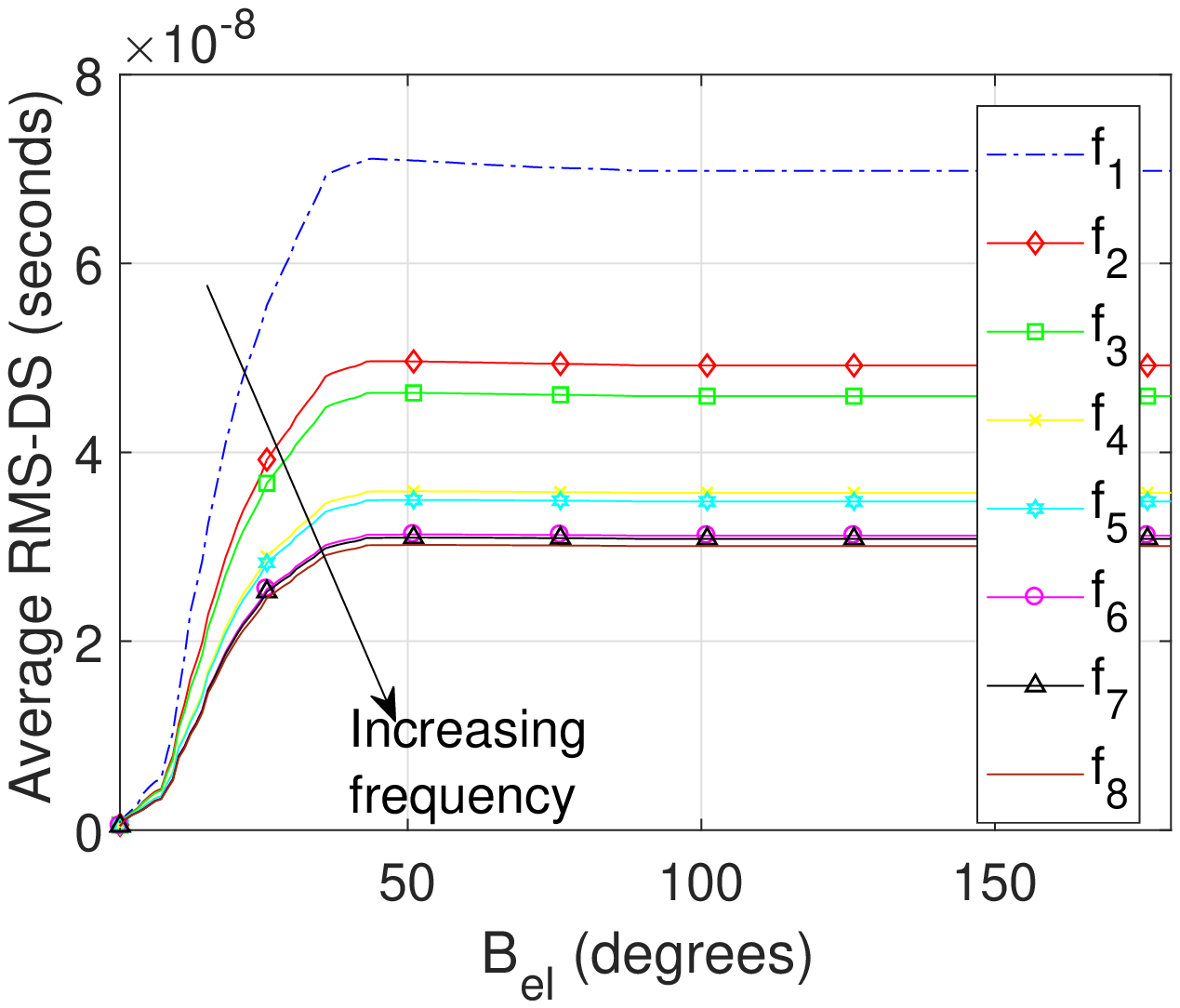}} 
        \caption{Elevation plane.}
    \end{subfigure}
    \caption{\textbf{Left:} Average AOA RMS-AS at the UE; \textbf{Middle:} Average AOD RMS-AS at the BS; and \textbf{Right:} Average RMS-DS at the UE for LOS trajectory. (a) illustrates the results in azimuthal plane while (b) is for the elevation plane.}\vspace{-0.45cm}
    \label{AS_AOD}
\end{figure*}
\subsection{Directional Angular Spread and Delay Spread}
In this subsection, we evaluate the channel time and angular dispersive properties, and analyze the RMS-DS and RMS-AS of the MPCs. For estimation of the AOA/AOD directional AS, we calculated the AOA/AOD RMS-AS as a function of BeW from the most dominant path. That is, we apply a continuous BA technique between the BS and the UE for detecting the most dominant MPC. In other words, we continuously latch on to the strongest MPC. The most dominant MPC's angle obtained from the BA is treated to be the best transmit and receive direction, which in turn corresponds to the boresight angle of the directional transmission/reception antennas at the BS and the UE, respectively.

We then calculate the RMS-AS by using the powers and angles of all MPCs within the BeW range. We assume the $B_{\rm el}$ and $B_{\rm az}$ as the BeW in the elevation and azimuthal plane, respectively. Then, the mean AoA evaluated at the vehicle location $n$ is calculated as follows:
\vspace{-0.2cm}
\begin{equation}\label{Mean_Formula}
\bar{\phi} (n) =\frac{\sum_{k=1}^{\hat{K}_{\rm BeW}(t)} \psi_k(n) |\gamma_k(n)|^2}{\sum_{k=1}^{\hat{K}_{\rm BeW}(t)} |\gamma_k(n)|^2},
\end{equation}
where $\hat{K}_{\rm BeW}(t)$ corresponds to the number of MPCs within the BeW considered. Here, $\psi_k(n)$ and $\gamma_k(n)$ correspond to the AOA and AnP term of the $k^{th}$ MPC at point $n$, respectively. The BeW considered can be either $B_{\rm el}$ or $B_{\rm az}$ depending on the plane considered. Subsequently, the RMS-AS can be obtained as follows:
\begin{equation}\label{RMS_Formula}
{\phi}^{\rm RMS} (n) =\sqrt{\frac{\sum_{k=1}^{\hat{K}_{\rm BeW}(t)} \left( \psi_k(n) - \bar{\phi} (n)\right)^2|\gamma_k(n)|^2}{\sum_{k=1}^{\hat{K}_{\rm BeW}(t)} |\gamma_k(n)|^2}}~.
\end{equation}
While calculating the AS in the azimuthal plane we assume that the entire domain of the elevation plane is considered, and vice-versa. Similarly, one can modify the above formulas to calculate the AS for the AOD term, and a similar concept is used to calculate the RMS-DS. 

\section{Simulation Results}
In this section, we present results on temporal/spatial V2X channel statistics, as well as the impact of beamforming and blockage on such statistics, after processing the data obtained from RT simulations. Unless mentioned otherwise, all the results presented are averaged over all the possible vehicle locations in the LOS and NLOS trajectory, respectively. 



\subsection{AOA/AOD RMS Angular Spread}
Results in Fig.~\ref{AS_AOD} (left and middle) illustrate the average AOA and AOD RMS-AS at the UE and BS, respectively, as a function of the BeW in the azimuthal and elevation planes. The plot is specific to the LOS trajectory. The AS effect is somewhat similar across the frequency bands; however, it diminishes with the increase in frequency. This is obvious because the PL increases with the frequency as seen in Fig.~\ref{Avg_PL_Multiple_Bounce}. As the BeW increases, more and more MPCs fall within the considered BeW range resulting in a general trend of monotonic increase in AS. The increase will not be significant when the strength of the MPC is negligible. Both in the AoA and AoD domain, the AS in the azimuthal plane is significant relative to the elevation plane. These plots suggest that the RMS-AS is significantly lesser at the mmWave bands relative to the sub-6 GHz bands, assuming that similar directional transmission/reception are considered across all frequency bands. 
The \emph{significance of these results} is that they give us a sense of appropriate BeW to be selected at the BS/UE in both the azimuthal and elevation planes to capture the significant MPCs based on a particular frequency band.


\vspace{-0.15cm}
\subsection{RMS Delay Spread}
Fig.~\ref{AS_AOD} (right) shows the RMS-DS as a function of the BeW in the elevation and azimuthal planes. It is observed that the RMS-DS becomes smaller with the increase in the frequency, hinting at a weaker strength of the MPCs as the frequency increases. The large RMS-DS is due to multiple reflections and scattering which causes large delays with respect to the dominant signals. As observed, the DS in the azimuthal and elevation plane look similar. The plots suggest that having a limited BeW at the BS and the UE would still capture significant MPCs. These results were obtained with equal antenna gain for all spatial directions. In the next subsection, we discuss the effect of antenna gain on directional transmission/reception, in particular for the $f_6$ band.


\begin{figure}[t!]
\centerline{\includegraphics[scale=0.5]{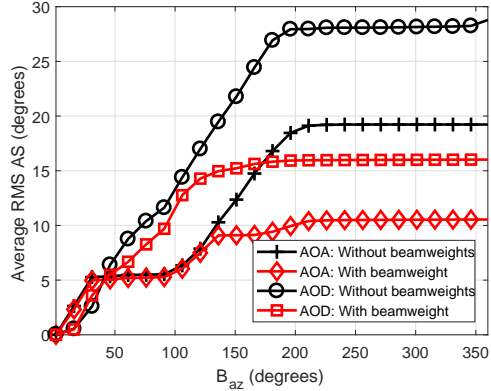}}
\caption{Effect of antenna gain on the RMS-AS in the azimuthal plane of AOA and AOD domain for LOS trajectory at $f_6$ band.}\vspace{-0.35cm}
\label{RMS_AS_Beamweight}
\end{figure}

\subsection{Effect of Antenna Gain}\label{Antenna_Gain}
The results presented so far are obtained for equal antenna gain for all spatial directions. Since no spatial selectivity is considered, it captures all existing MPCs between transmitter and receiver with unity gain. To model the directional transmission/reception with antenna gain, we take into account the antenna model \cite{Antenna_Model} which relates the antenna gain and BeW by a Gaussian main lobe profile which is given by:
\begin{eqnarray}\vspace{-0.05cm}
\begin{aligned}
G_{\rm dB}(\Phi_{\rm az},\Phi_{\rm el}) &=10 \log \left(\frac{16 \pi}{6.76 w_{\rm 3dB}^{\Phi_{\rm az}} w_{\rm 3dB}^{\Phi_{\rm el}}} \right)\\& -12\left(\frac{\Phi_{\rm az}}{w_{\rm 3dB}^{\Phi_{\rm az}}} \right)^2-12\left(\frac{\Phi_{\rm el}}{w_{\rm 3dB}^{\Phi_{\rm el}}} \right)^2~,
\end{aligned}\vspace{-0.05cm}
\label{Antenna_Gain_Formula}
\end{eqnarray}
where $\Phi_{\rm az}$ and $\Phi_{\rm el}$ are defined as offsets between the main lobe direction and the azimuth angle and elevation angle, respectively, while $w_{\rm 3dB}^{\Phi_{\rm az}}$ and $w_{\rm 3dB}^{\Phi_{\rm el}}$ are the azimuth and elevation half-power beamwidths (HPBW), respectively. We investigate the effect of directional transmission/reception antenna gain only for the mmWave bands.

We explore the effect of the antenna gain on the RMS-AS and RMS-DS in the azimuthal plane for the LOS trajectory with fixed elevation angle (2D environment). Based on the results from previous subsection, we set the elevation HPBW to ${30}^{\circ}$. We calculate the RMS-DS and RMS-AS by incorporating the beam-weights into \eqref{Mean_Formula}-\eqref{RMS_Formula}. Fig. \ref{RMS_AS_Beamweight} plots the RMS-AS in the azimuthal plane of AOA and AOD domain for $f_6$ (28~GHz) with and without the directional antenna gain (referred as beamweight). As seen, the directional transmission and reception with beamweigths reduces the RMS-AS significantly. The main reason for this behavior is due to the fast decay of the beamforming weights from the boresight of the beam direction, which (when compared to equal antenna gain) suppresses the MPCs departing/arriving at those corresponding angular directions. The RMS-DS plot and analysis is not added due to space constraint. However, we note that the trend was similar to the RMS-DS plot.

\begin{figure}[t!]
\centerline{\includegraphics[width=12cm,height=6cm]{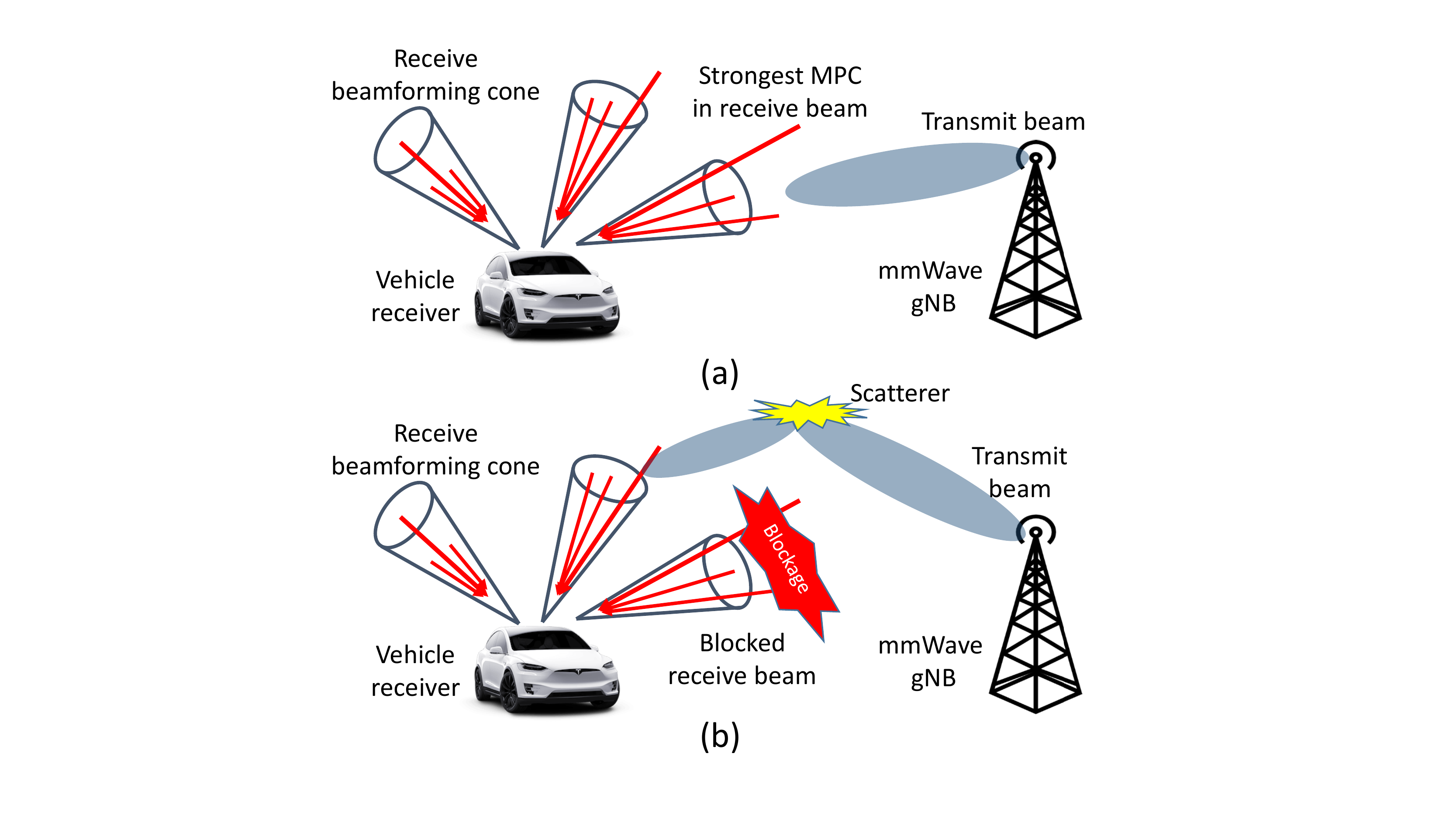}}\vspace{-0.45cm}
\caption{Illustration of cone blockage in spatial domain. (a) The MPCs are assumed to arrive at the vehicle within distinct cones each of which may be individually subject to blockage. (b) If a particular cone gets blocked, the receiver locks to the strongest MPC outside of this blocked cone}\vspace{-0.6cm}
\label{3D_Cone_Fig}
\end{figure}

\begin{figure}[b!]
\centerline{\includegraphics[scale=0.335]{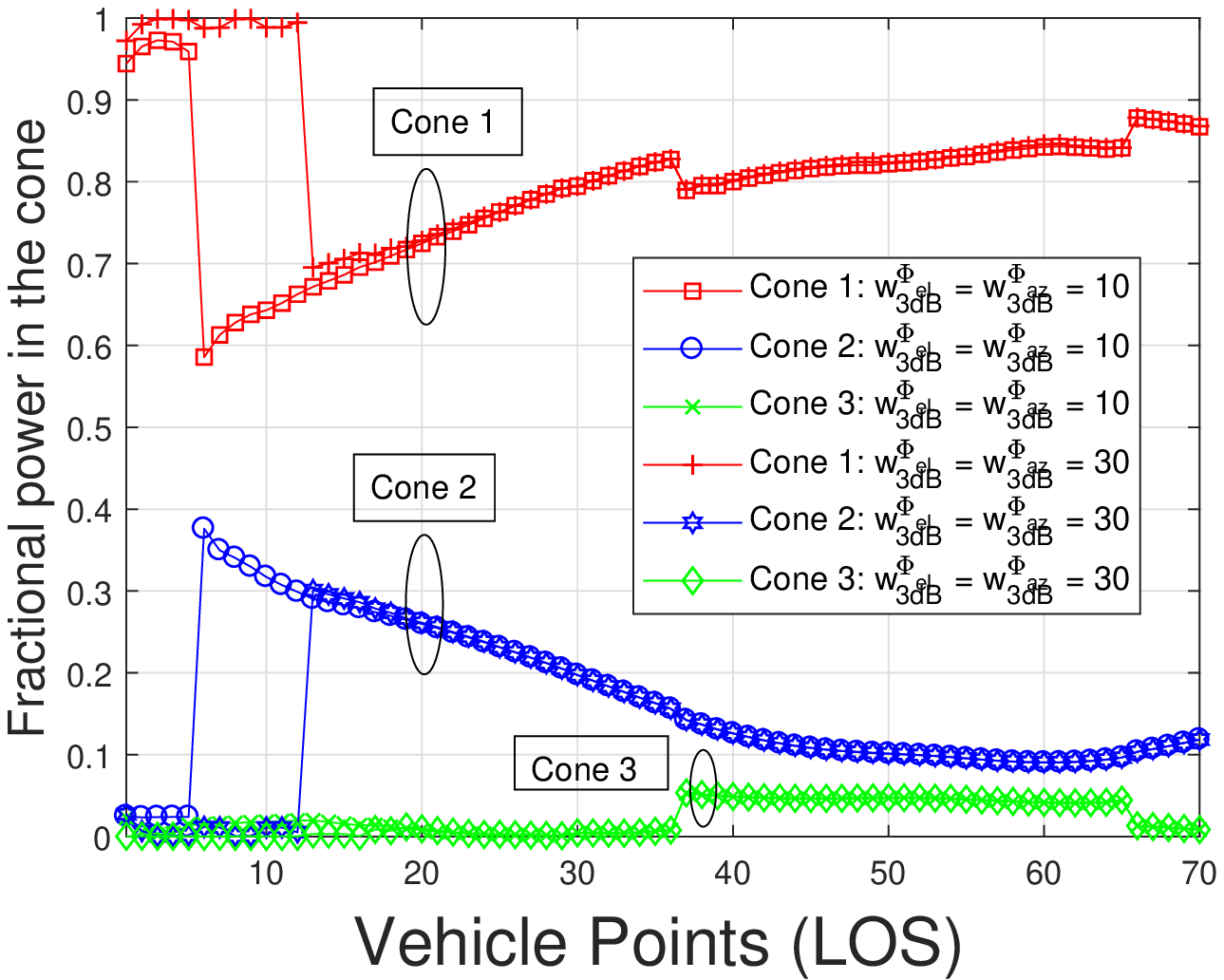}\hspace{-0.4cm}\includegraphics[scale=0.335]{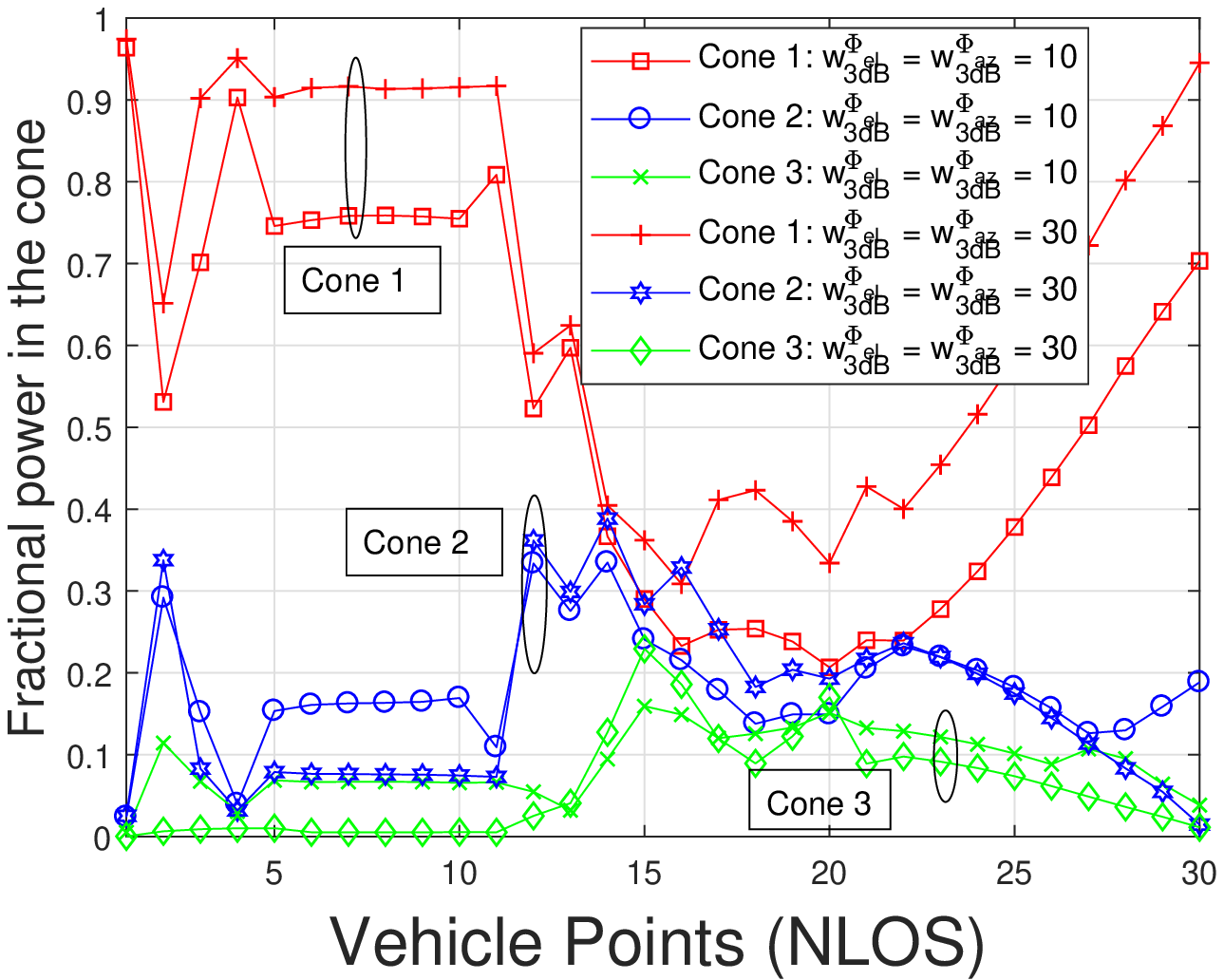}}
           \caption{Fractional MPC power in cones with fields of view of $10^{\circ}$ and $30^{\circ}$, at different points on the vehicle trajectory. \textbf{Left:} LOS trajectory and \textbf{Right:} NLOS trajectory.}\vspace{-0.1cm}
           \label{Fractional_Power}
\end{figure}

\begin{figure*}[t!]
    \centering
    \begin{subfigure}[t]{0.5\textwidth}
			\centerline{\includegraphics[scale=0.425]{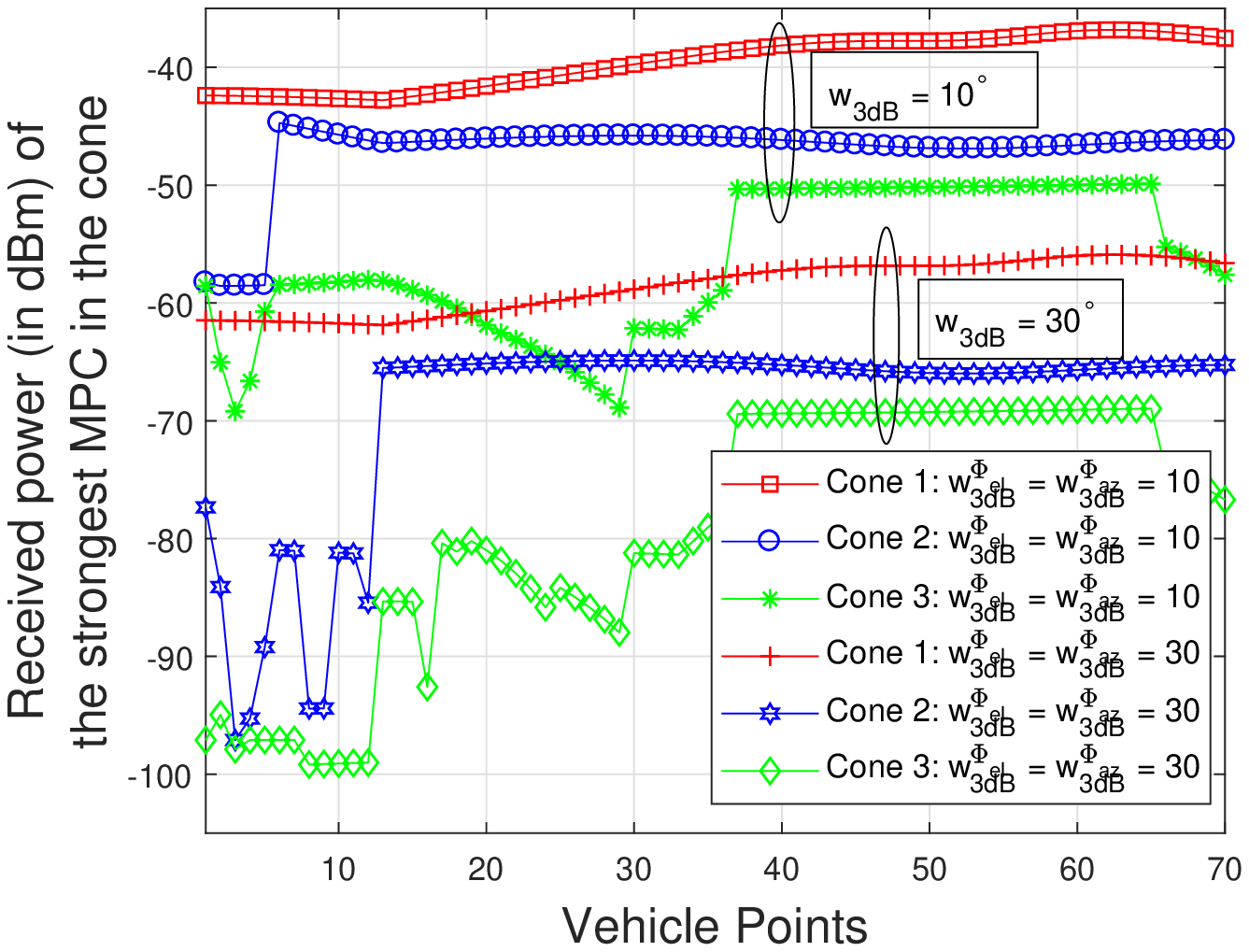}\hspace{-0.4cm}\includegraphics[scale=0.425]{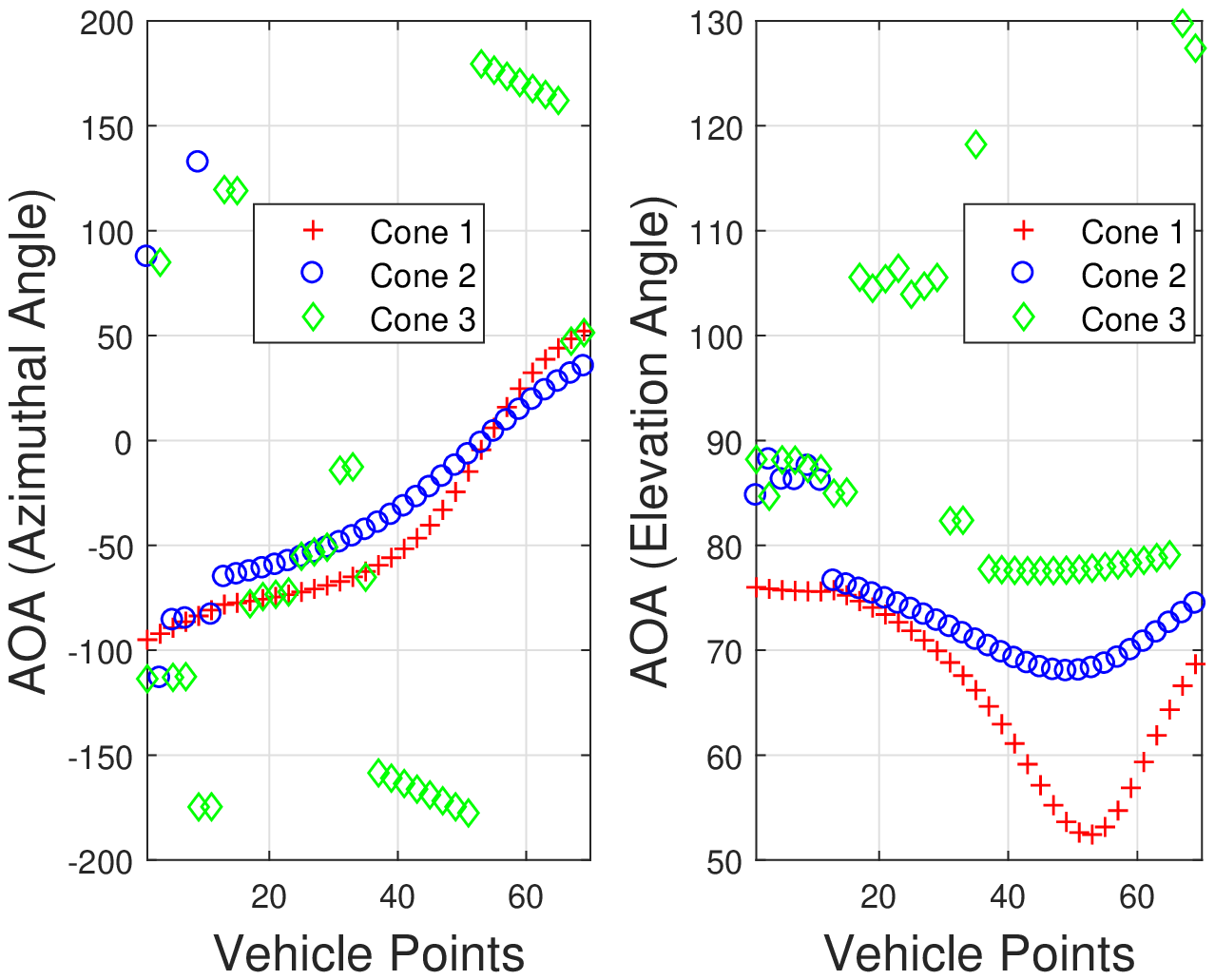}\hspace{-0.4cm}\includegraphics[scale=0.425]{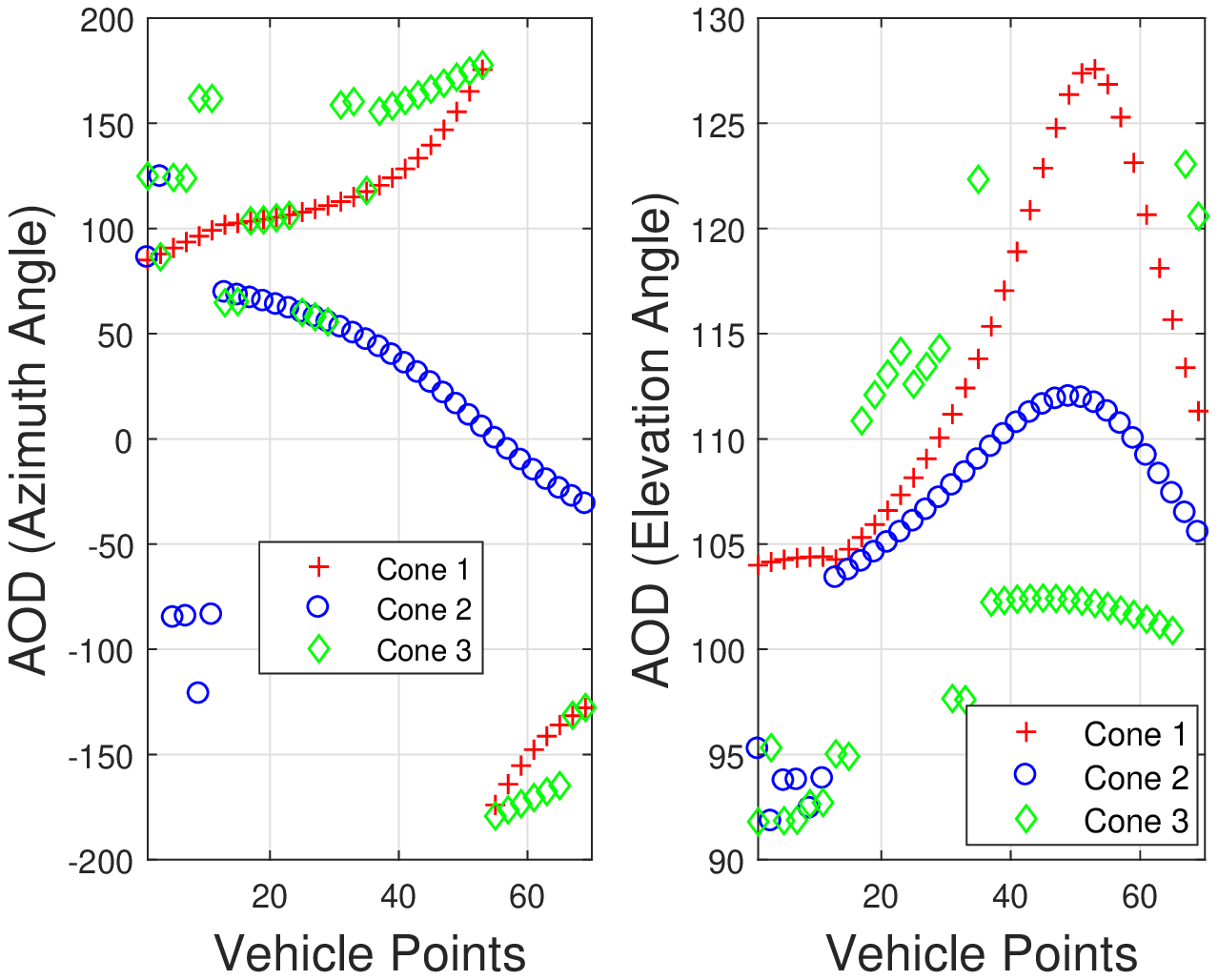}}
           \caption{LOS trajectory.}
    \end{subfigure}%
        \vspace{-0.05cm}
    \begin{subfigure}[t]{0.5\textwidth}
			\centerline{\includegraphics[scale=0.425]{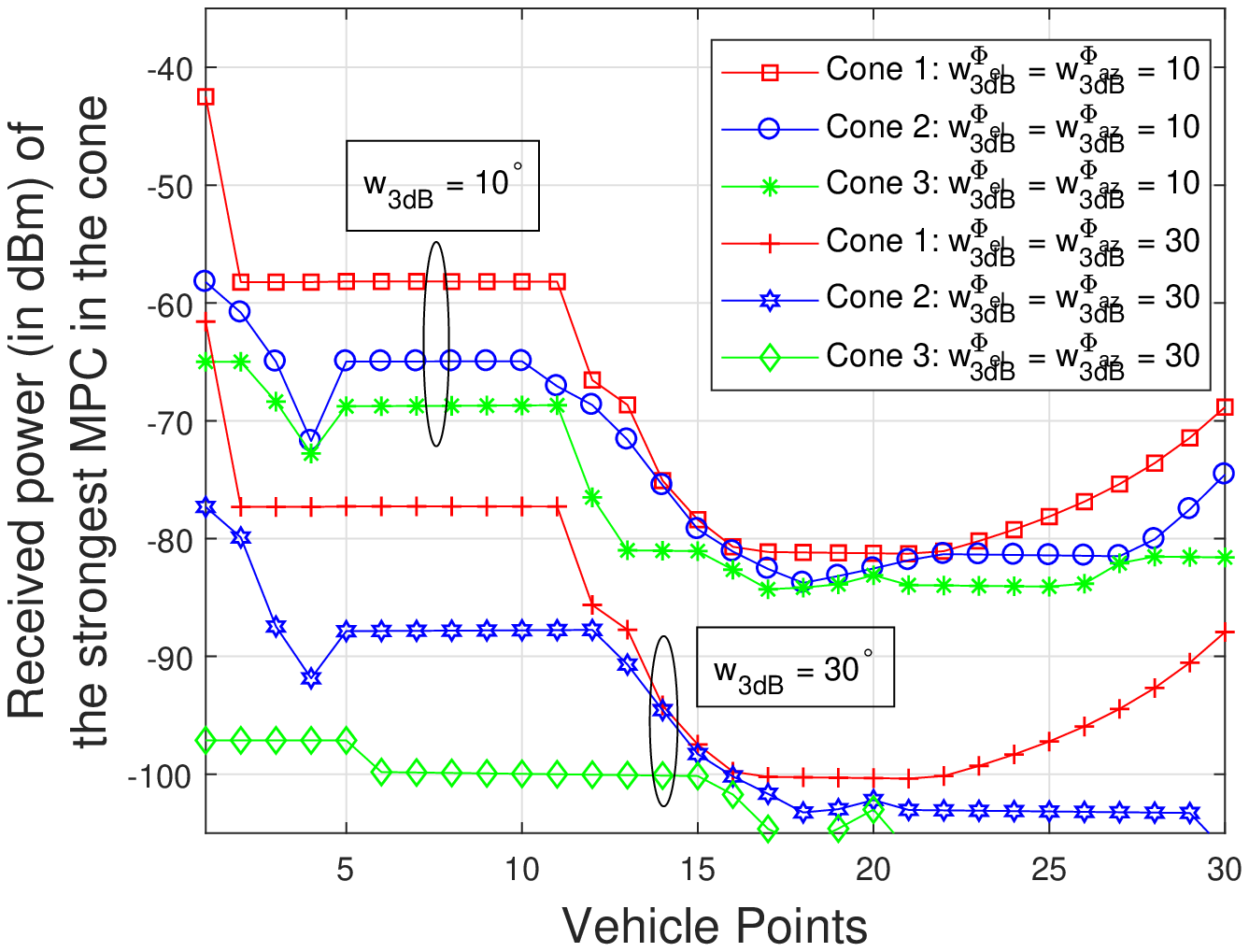}\hspace{-0.4cm}\includegraphics[scale=0.425]{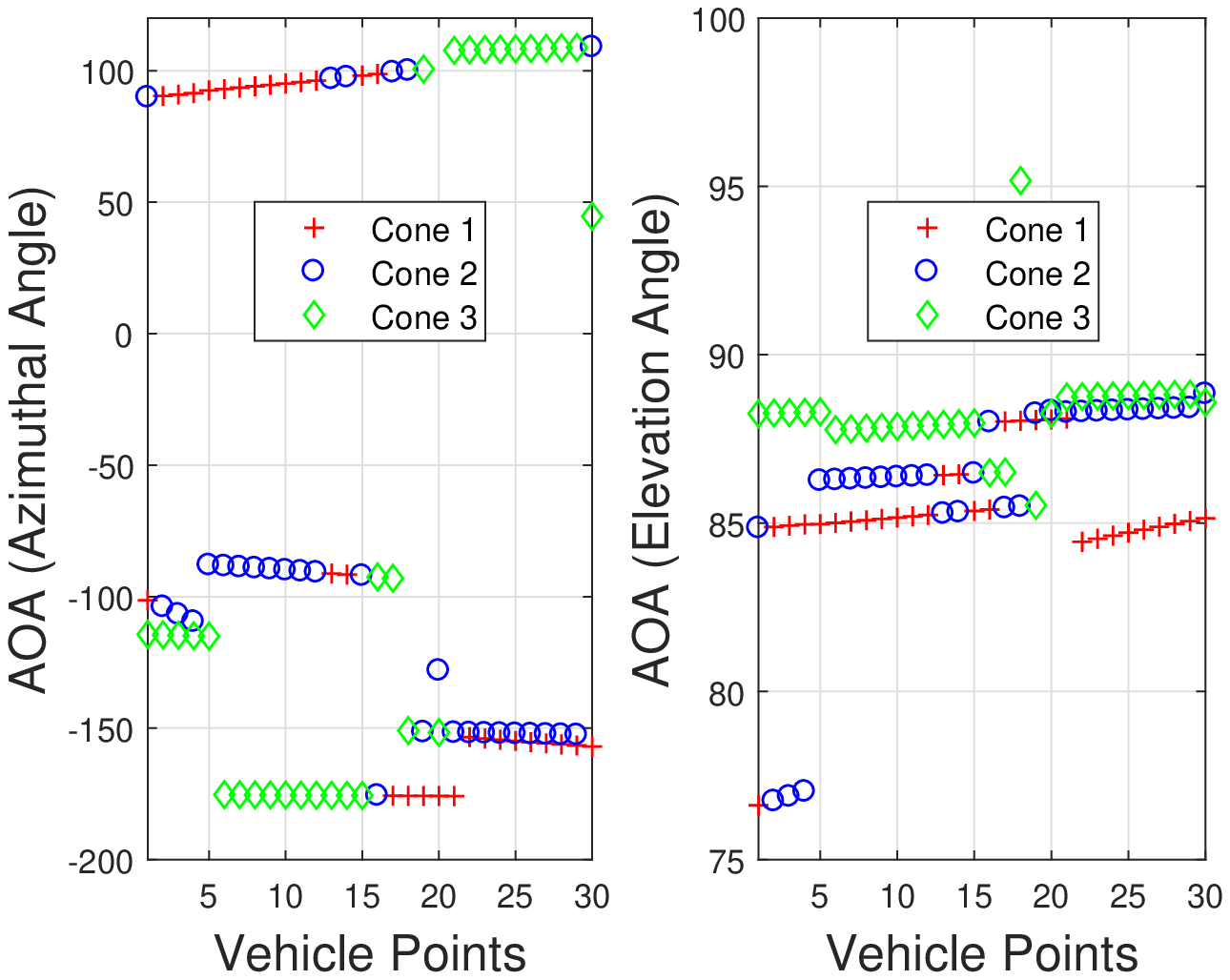}\hspace{-0.4cm}\includegraphics[scale=0.425]{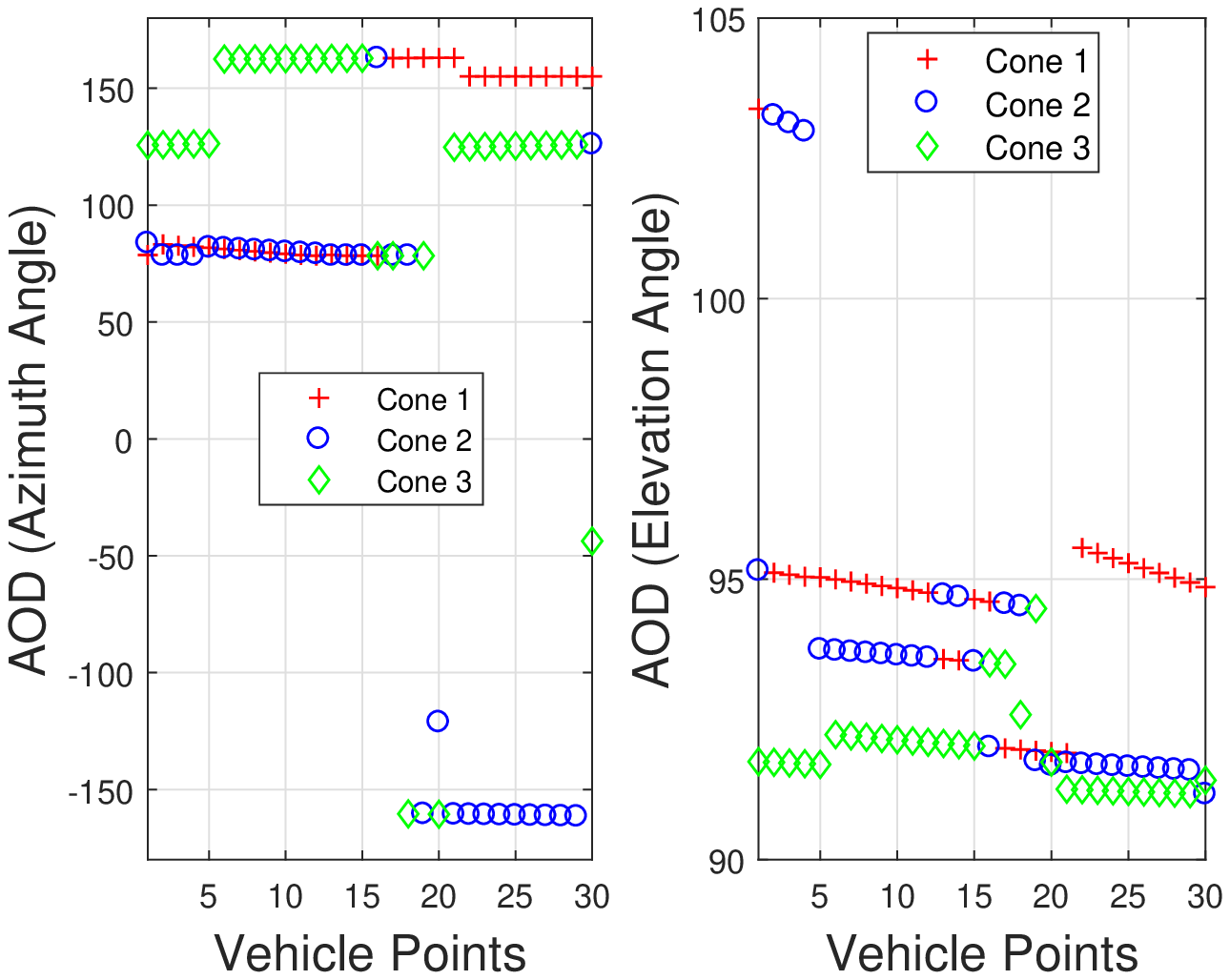}}
           \caption{NLOS trajectory.}
           \end{subfigure}
           \caption{\textbf{Left:} Received power of the strongest MPC in cone at band $f_6$. \textbf{Middle:} AOA characteristics with $w_{3dB}^{\Phi_{el}} = w_{3dB}^{\Phi_{az}} = 30^{\circ}$. \textbf{Right:} AOD characteristics with $w_{3dB}^{\Phi_{el}} = w_{3dB}^{\Phi_{az}}$ = $30^{\circ}$ at different vehicle locations in the trajectory.}\vspace{-0.35cm}
           \label{RP_Spatial_Charactersrtics}
\end{figure*}

\subsection{Spatial Blockage}
In this subsection, we investigate the effect of blockage in the spatial domain for the considered vehicle trajectories. In particular, we consider that the receiver beamforming captures the MPCs within a cone, which may be subject to blockage as illustrated in Fig.~\ref{3D_Cone_Fig}. 
When a receive beam is blocked (we assume whole cone is blocked), it is necessary to find alternative beam directions (cones non-overlapping with the blocked cone) to maintain the communication link.
In here, we assume that the spatial information from a lower band $f_5$ is available and is used to perform BA at band $f_6$. To construct a cone, we latch onto the strongest MPC and construct a 3D cone around it. The width of the 3D cone is dictated by the choice of HPBW $w_{\rm 3dB}^{\Phi_{\rm az}}$ and $w_{\rm 3dB}^{\Phi_{\rm el}}$. As seen in (\ref{Antenna_Gain_Formula}), lower the HPBWs, higher the beam-forming gain and lower the coverage in spatial domain, and vice versa. The angle of the strongest MPC corresponds to the bore sight angle of the cone constructed. Similarly, other non-overlapping cones are constructed by taking the strongest MPC outside the existing cones. For the simulations, we choose the HPBW $w_{\rm 3dB}^{\Phi_{\rm az}}$ and $w_{\rm 3dB}^{\Phi_{\rm el}}$ both at the BS and the UE to be $10^{\circ}$ and $30^{\circ}$. 

Fig. \ref{Fractional_Power} plots the fractional power of the cones across the vehicle's trajectory in Fig.~\ref{Ray_Tracing1} for different selection of the HPBWs. The main objective of this study is to see the power concentration in angular domain, over the trajectory of the vehicle. It is evident that almost all the power is concentrated around first three (non-overlapping) cones that lie apart in the angular domain. For the LOS trajectory, Cone~1 involves the direct path and carries most of the power, followed by Cone 2 and 3. If any of the cones are blocked, then the communication can fall back to other cones which are segregated in the spatial domain assuming these cones are not blocked. 

For the LOS trajectory in Fig. \ref{Fractional_Power}, note the similarity in fractional power of the cones at different HPBWs except at vehicle locations 5 through 12. The overlap in Cone 1 is because most of the MPCs that depart from the BS and arrive at the UE falls within the HPBW $10^{\circ}$. In other words, the MPCs are closely clustered around the LOS path in the spatial domain. However, at point 5 through 12 the MPCs are loosely clustered and they fall under Cone 2, resulting in a different power concentration that depends on the choice of HPBWs. The same phenomenon is witnessed in the NLOS trajectory because the MPCs are loosely clustered in the spatial domain.

Fig. \ref{RP_Spatial_Charactersrtics} (left) plots the received power of the strongest MPC in the cone for the LOS and NLOS trajectories. Fig.~\ref{RP_Spatial_Charactersrtics} (middle and right) plots the angular characteristics in the AOA and AOD domain, respectively for different vehicle locations, again for LOS and NLOS scenarios. These figures combined together gives an insight about the strength of the strongest received MPCs after a blockage, and the angular trajectory at different locations for the considered vehicle trajectories. For LOS trajectory, the strongest MPC corresponds to the direct path and falls in Cone~1. The angular characteristics of the direct path often changes smoothly and is relatively easy to track. However, the angular characteristics of strongest MPC of the secondary cones vary rapidly as seen in Fig.~\ref{RP_Spatial_Charactersrtics} (middle, right). For the NLOS trajectory, the transitions are not smooth and are often difficult to track.

With lower HPBW, the received power of the strongest MPC will be high due to higher directivity gain. However, the chances of blockage or misalignment of the entire cone is higher compared to higher HPBW. Thus, the HPBW should be chosen based on the trade-off between the possibility of blockage (priori information) and received strength. The use of priori information will be explored in our future work.



\section{Conclusion}
In this work, we used RT simulation data to investigate the temporal and angular characteristics of sub-6 GHz and mmWave bands for a vehicular trajectory in an urban setting. Impact of beam width, beam shape, and spatial blockages on such characteristics are also explored. The study of channel characteristics across different bands helps one to better exploit the OOB information for a robust and efficient communication, and to design effective beam tracking techniques for mobile vehicular channels.


\bibliographystyle{IEEEtran}
\bibliography{bib_ICC}



\end{document}